# ORIGIN OF THE HYDROGEN INVOLVED IN IRON CORROSION UNDER IRRADIATION

S. Lapuerta[a,b], N. Millard-Pinard[a], N. Moncoffre[a], N. Bérerd[a], H. Jaffrezic[a], G. Brunel[a],

D. Crusset[b], Th. Mennecart[c],

[a] Institut de Physique Nucléaire de Lyon, 4, rue Enrico Fermi, 69622 Villeurbanne cedex, France,

[b] ANDRA, Parc de la Croix Blanche 1-7 rue Jean Monnet, F-92298 Châtenay-Malabry Cedex, France

[c] CNRS-CERI, 3A rue de la Férollerie, 45071 Orleans Cedex, France

*Corresponding author:* N. Millard-Pinard

Email : millard@ipnl.in2p3.fr

Phone : +33 4 72 44 80 62

Fax : +33 4 72 44 80 04

**Abstract**

In the perspective of long term geological storage, high level nuclear wastes will be overpacked in low carbon steel containers. In that context, we have studied the influence of oxygen dissolved in water on iron corrosion. Therefore, leaching experiments were performed in desaerated $D_2O$ and in aerated $H_2O$ and a kinetic study of iron corrosion under proton irradiation was lead in aqueous media with two different dissolved oxygen concentrations. The leaching experiments underline the major role of dissolved oxygen in oxydoreduction reactions which take place as far as iron is in contact with water. But the kinetic study of iron corrosion under irradiation put in evidence the balance between the oxydoreduction reactions and the corrosion rate induced by radicals species generated by water radiolysis.

In addition, to check if, in the atomic % concentration range, hydrogen diffuse from the air/Fe interface through the foil an irradiation experiment was performed in argon. It proved that no hydrogen permeation occurs at a concentration level of the atomic percent.

# 1. Introduction

In the perspective of long term geological disposal, high level nuclear wastes will be overpacked in low carbon steel containers. In a recent paper [1], the influence of water radiolysis on iron corrosion was studied by using 12 MeV protons with a 30 nA intensity. Experiments were realised with 2 different water media : aerated deionised water ($H_2O$) and heavy water ($D_2O$) packaged under argon. It was shown that after a two hour irradiation time, the iron loss on the sample irradiated in aerated water was larger than 60 nm thick but was negligible in case of desaerated $D_2O$ medium. It was suggested that, as observed by Burns [2], the enhancement of corrosion in aerated water was due to $O_2$ dissolved species. Another unexpected result is the similar hydrogen profiles observed in samples irradiated in desaerated $D_2O$ and aerated $H_2O$. Such results could be attributed to the well known D/H isotopic exchange but it was also suggested that hydrogen, because of its high diffusivity, comes from the wet air and diffuses inside the iron foil.

The aims of the work presented in this paper are respectively:

- to study the impact of dissolved oxygen in oxydoreduction reactions which take place at the iron water interface. For that purpose, two experiments were performed:

    i) The comparison between the iron surface composition evolution after leaching experiments in desaerated $D_2O$ and in aerated $H_2O$.

    ii) The study of the iron corrosion kinetics under proton irradiation in aqueous media with two different dissolved oxygen concentrations.

- to answer the relevant question: do hydrogen ions, produced in wet air by irradiation, migrate through the whole sample thickness? In order to avoid radiolysis effects which could hide migration of hydrogen species, an irradiation experiment was performed in which the irradiation cell was filled with argon instead of water.

**2. Influence of oxygen dissolved species in water on iron aqueous corrosion**

*2.1. Leaching experiments in aerated $H_2O$ and in desaerated $D_2O$*

Leaching experiments were performed at 40°C during 150 hours in closed vessel on pure iron (99.995 %) samples (250 µm thick). To avoid isotopic exchange [3], the sample was analysed by ion beam techniques using the 4 MV Van de Graaff accelerator of the "Institut de Physique Nucléaire" of Lyon (IPNL), just after the end of the leaching procedure. The hydrogen profiles were measured using Elastic Recoil Detection Analysis (ERDA) induced by 1.7 MeV alpha particles. The deuterium profiling was performed by Nuclear Reaction Analysis (NRA) using the $D(^3He,\alpha)p$ nuclear reaction. In this reaction, the difference in energy between the outgoing channel and the entrance one is 18.35 MeV. Thus, the energy of the emitted α particles is much larger than the $^3He$ incident energy. In order to favour the deuterium concentration measurement at the sample surface, a 15° glancing angle was used. The experimental set up is shown in Fig. 1. The α energy distribution was measured at a 30° detection angle, while the scattered $^3He$ ions are stopped in a 6.5 µm thick mylar absorber. Therefore, the α detection energy resolution is 50 keV. A 800 keV incident energy was chosen because, in the 600 to 800 keV energy range, the reaction cross section can be considered as constant. This approximation is relevant for two reasons:

    i) W.E. Kunz [4] has shown that in the 100-800 keV incident energy range, the α particle angular distribution is isotropic.

    ii) The reaction cross section [5] varies smoothly in the 600-800 keV energy range.

A deuterium standard elaborated by ion implantation in a $Si_3N_4/Si$ sample was used to normalize the concentration measurements [6].

It must be noted that NRA and ERDA are elemental analysis but other characterisation techniques have identified hydroxyl species [6]. As shown in Fig. 2, at a 40 nm iron depth, the order of magnitude of the [H]/[D] ratio is 150. Actually, the isotopic effect on the slowing down of $D_2O$ reaction velocity [7] compared to $H_2O$ can explain a maximum [H]/[D] ratio value equal to two. Thus, the very low deuterium concentration corresponding to desaerated $D_2O$ leaching, compared to the high hydrogen concentration value in case of leaching in aerated $H_2O$, emphasizes the major role of dissolved oxygen on the reaction at the Fe/water interface.

These oxydoreduction reactions are:

$$2\ Fe \rightarrow 2\ Fe^{2+} + 4\ e^- ,  \quad (1)$$

$$O_2 + 4e^- + 2\ H_2O \rightarrow 4\ HO^- ,  \quad (2)$$

$$2\ Fe + O_2 + 2\ H_2O \rightarrow 2\ Fe^{2+} + 4\ HO^- .  \quad (3)=(1)+(2)$$

These reactions also occurred for iron corrosion in wet air [8].

The following part deals with the influence of dissolved oxygen concentration on aqueous iron corrosion under proton irradiation.

*2.2 Study of the iron aqueous corrosion kinetics under proton irradiation*

The iron corrosion kinetics under proton irradiation was performed in two aqueous media: aerated $H_2O$ in which dissolved oxygen is in equilibrium with air and $H_2O$ under argon bubbling called $H_2O(Ar)$ medium. The dissolved oxygen concentrations are respectively equal to $2.5 \times 10^{-4}$ mol $L^{-1}$ in aerated $H_2O$ and $2.3 \times 10^{-6}$ mol $L^{-1}$ in $H_2O(Ar)$ medium. They have been controlled by the voltamperometry technique.

Let us remind the experimental conditions which were already described in details in [1]. The pure iron samples are europium implanted on the polished face at a 800 keV energy and a $5 \times 10^{15}$ at. cm$^{-2}$ fluence. This europium tracer is used to measure the iron loss thickness in the corrosion process. In order to ensure, that all the studied samples present the same surface state before irradiation, a pre-irradiation procedure has been defined. The iron foil is the entrance window of the irradiation cell (Fig. 3), its polished side being put in contact with $H_2O$ during 12 hours. Then, the cell water is changed. After this procedure, a sample called "blank" is used as reference and the other samples are irradiated.

The irradiation experiments were performed with the CERI (Centre d'Etudes et de Recherches par Irradiation) cyclotron at Orleans, which delivers a 12 MeV proton beam. The irradiation set up is presented in Fig. 3. Irradiations were carried out with a 30 nA beam intensity, the irradiation time varying from 5 to 75 minutes.

Ion beam analysis (Rutherford Backscattering Spectrometry (RBS) and ERDA) were performed using 1.7 MeV α particles. The SIMNRA program was used to simulate the energy spectra obtained both by RBS and ERDA so as to determine the atomic concentration profiles of europium, oxygen and hydrogen. For a given sample analysed by RBS and ERDA, the analysis was done by iterations, taking always into account, in the simulations files, the europium, oxygen and hydrogen concentrations.

The iron loss thickness deduced from the europium signal vanishing is presented in Fig. 4. The iron loss velocities, calculated from a linear regression, are respectively 1.7 nm min$^{-1}$ in case of aerated water and 1.1 nm min$^{-1}$ in case of $H_2O(Ar)$ medium. Although the dissolved oxygen concentration ratio in respectively $H_2O$ and $H_2O(Ar)$ is 100, the iron corrosion velocity ratio is less than two. This result can be explained by radiolysis effects occurring under proton irradiation even in desaerated medium. The crucial role on the corrosion process of the radical species formed near the sample surface was also observed in reference [9].

Indeed, in desaerated water, the radical and molecular species due to water radiolysis induced the following reactions:

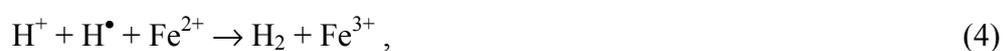

$H^+ + H^\bullet + Fe^{2+} \rightarrow H_2 + Fe^{3+}$ , (4)

$$HO^\bullet + Fe^{2+} \rightarrow OH^- + Fe^{3+} , \tag{5}$$

$$H_2O_2 + Fe^{2+} \rightarrow Fe^{3+} + HO^\bullet + HO^- . \tag{6}$$

The global reaction is :

$$2\ H^+ + 2\ Fe^{2+} \rightarrow H_2 + 2\ Fe^{3+} . \tag{7}=(4)+(5)\ or\ =(4)+(6)$$

In aerated water, the reaction (4) does not occur but the reactions (8) and then (9) take place.

$$O_2 + H^\bullet \rightarrow HO_2^\bullet , \tag{8}$$

$$3\ H^+ + HO_2^\bullet + 3\ Fe^{2+} \rightarrow 3\ Fe^{3+} + 2\ H_2O . \tag{9}$$

Then the global reaction is :

$$O_2 + 4\ H^+ + 4\ Fe^{2+} \rightarrow 2\ H_2O + 4\ Fe^{3+}. \tag{10}=(8)+(9)+(5)\ or\ =(8)+(9)+(6)$$

The Fig. 5 displays a comparison of oxygen and hydrogen profiles obtained in case of a 45 minutes irradiation in two media ($2.5\times10^{-4}$ mol L$^{-1}$ in aerated H$_2$O and $2.3\times10^{-6}$ mol L$^{-1}$ in H$_2$O(Ar)). These irradiations lead respectively to 105 and 77 nm iron losses. At a 100 nm depth, the hydrogen concentration for both cases are similar (around 10 at. %), while the oxygen concentrations are significantly different.

In fact, the hydrogen yield observed at the iron surface of the H$_2$O(Ar) irradiated sample can be related to the hydrogen radical species formed in desaerated water radiolysis. These H$^\bullet$ radicals can either induce reactions (reaction (4)) or migrate in the iron sample. During their short mean life time (around $10^{-6}$ s), they can migrate in the iron foil subsurface (up to 100 nm) as the hydrogen diffusion coefficient is large and range between $10^{-4}$ and $10^{-8}$ cm$^2$ s$^{-1}$ at 25°C [10,11].

Another hydrogen origin can be the charged species induced by proton irradiation in the 4 mm thick humid air gap between the titanium window and the iron foil (Fig. 3). Considering the high hydrogen diffusion coefficient in iron, it was necessary to check this hypothesis. Moreover, it must be noted that at much lower concentration ($10^{-6}$ at. %) hydrogen permeation in iron has been measured [6].

**IV Hydrogen permeation test**

The irradiation experiments were performed using the 4 MV Van de Graaff accelerator of the IPNL which delivers a 3 MeV proton beam. As shown in Fig. 6, the proton beam is extracted from the beam line vacuum to the atmosphere by crossing a 5 μm thick havar window. The external proton beam enters the irradiation cell through the studied 10 μm thick iron foil. To avoid any radiolysis effect, the irradiation cell is not filled with water but with argon. During irradiation, the beam intensity was set to 10 nA, and the irradiation time was equal to 90 minutes. The Relative atmosphere Humidity (RH) in the 8 mm gap between the havar window and the iron foil is controlled with a Hygropalm humidity controller and fixed to 45 %. Wayne-Siek et al. [12] have studied the charged species created in wet atmosphere under irradiation around the same RH. They have shown that more than 88 % of these species are $H^+(H_2O)n$ clusters. It can be assumed that these clusters decompose at the iron surface allowing the hydrogen penetration.

After irradiation, ERDA analysis was performed on both interfaces: air/Fe and Fe/Ar. The results are presented in Fig. 7. The 16 at. % hydrogen concentration measured at the Fe/Ar interface corresponds to the sample surface contamination, while it decreases very rapidly to a very small value. At the air/Fe interface, the hydrogen concentration stands around 10 at. % up to 100 nm depth. This result shows that hydrogen diffusion from the air/Fe interface through the foil is not observed.

**V Conclusion**

In this paper, we have shown that aqueous iron corrosion under proton irradiation is increased for high oxygen dissolved concentrations. Corrosion even in desaerated water under irradiation has been evidenced. It is explained by the formation of H$^\bullet$ and HO$^\bullet$ radicals which is large even in desaerated water. These H$^\bullet$ radicals can either migrate in the iron subsurface or induce corrosion process.

The similarity of hydrogen profiles observed at the Fe/water interface for irradiation performed either in oxygen rich or oxygen poor media has been pointed out. The origin of hydrogen enrichment at the Fe/water interface has been explained by the water radiolysis and not by the hydrogen diffusion from the air/Fe interface through the iron foil.

**Acknowledgments**

The authors would like to thank Catherine Corbel and Noelle Chevarier for numerous and enlightening discussions. We also thank Eric Mendes for his very helpful contribution in this work.

Figure captions

Fig. 1 : NRA experimental set up used to profile deuterium at the sample surface

Fig. 2: Hydrogen and deuterium profiles deduced by ion beam analysis

Fig. 3 : Schematic representation of the irradiation set up for iron leaching under proton beam.

Fig. 4: Dissolved iron tickness as a function of irradiation time for $H_2O$ and $H_2O(Ar)$ media

Fig. 5: Oxygen and hydrogen concentration profiles for $H_2O$ and $H_2O(Ar)$ media

Fig. 6: Schematic representation of the set up for the irradiation experiments at IPNL

Fig. 7: Comparison between hydrogen concentration profiles for 45 min irradiated sample at both interfaces

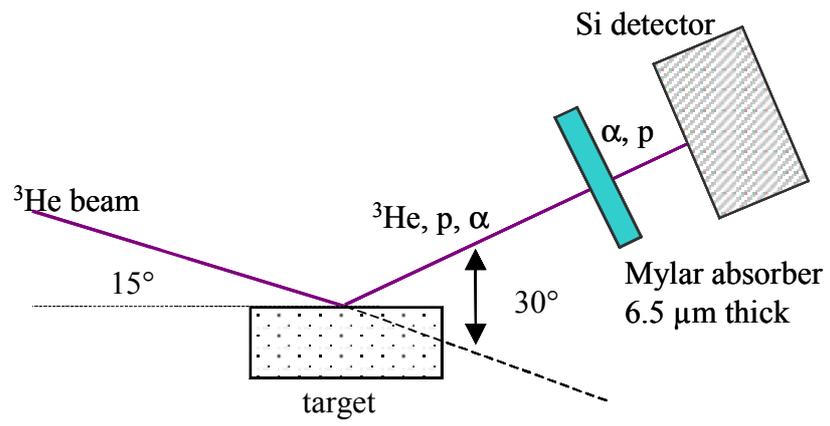

Figure 1

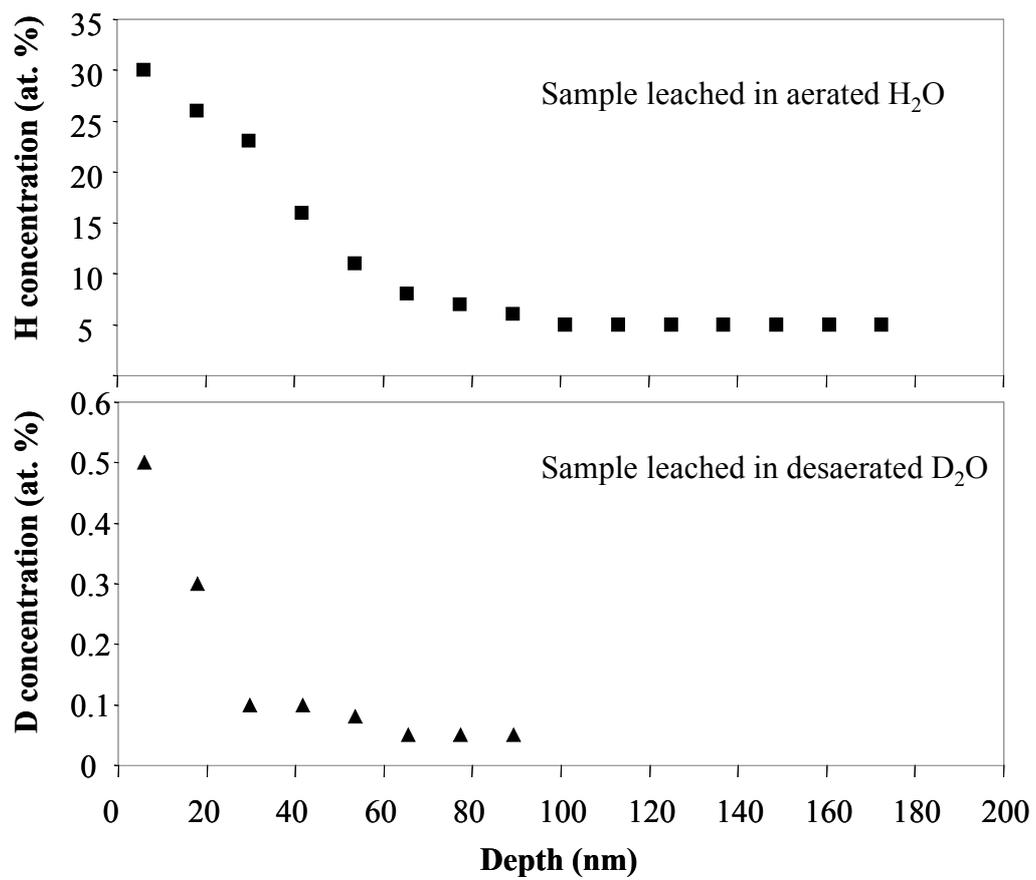

Figure 2

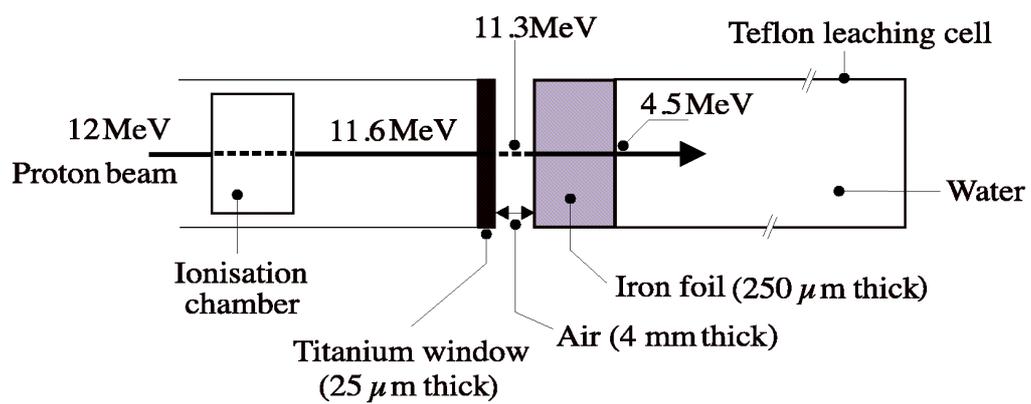

Figure 3

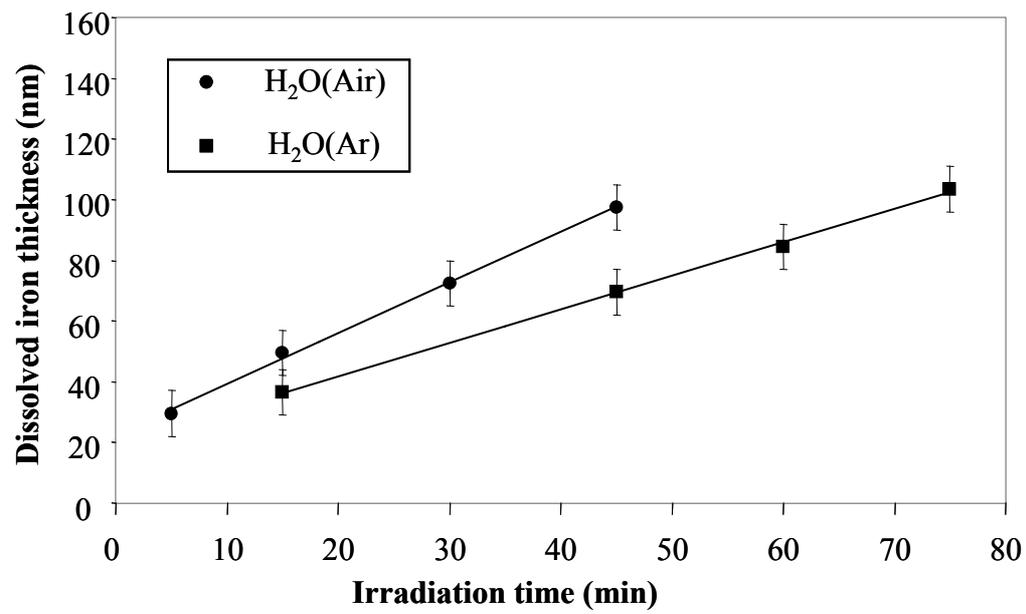

Figure 4

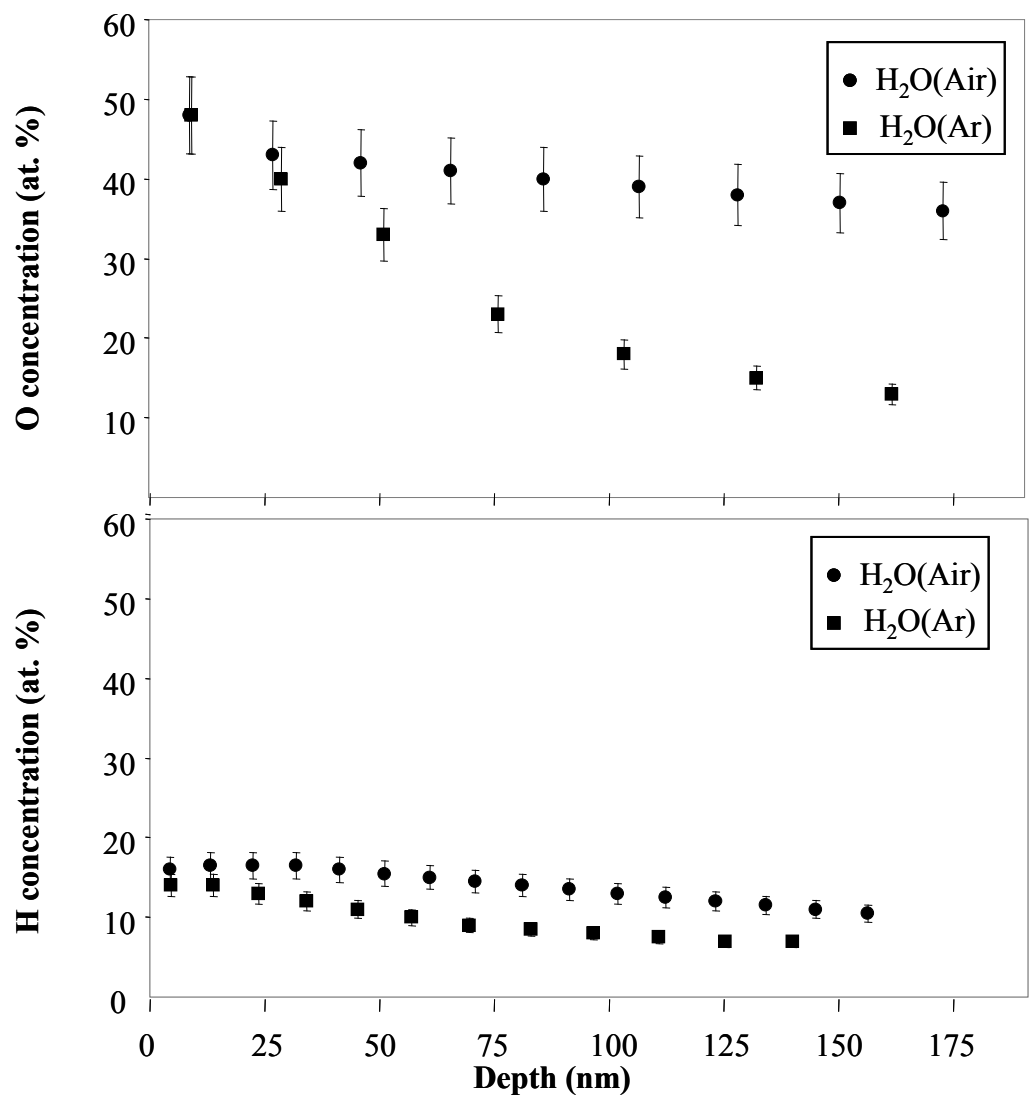

Figure 5

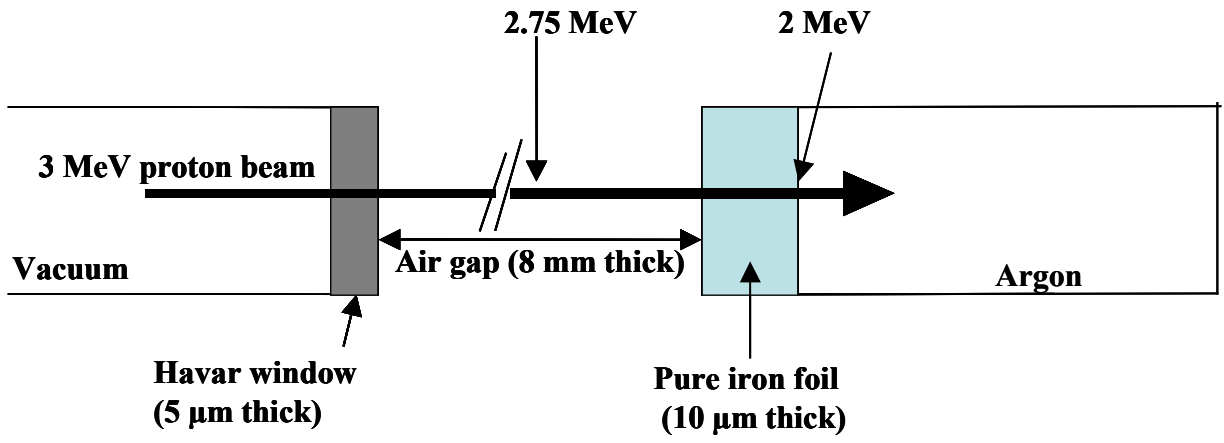

Figure 6

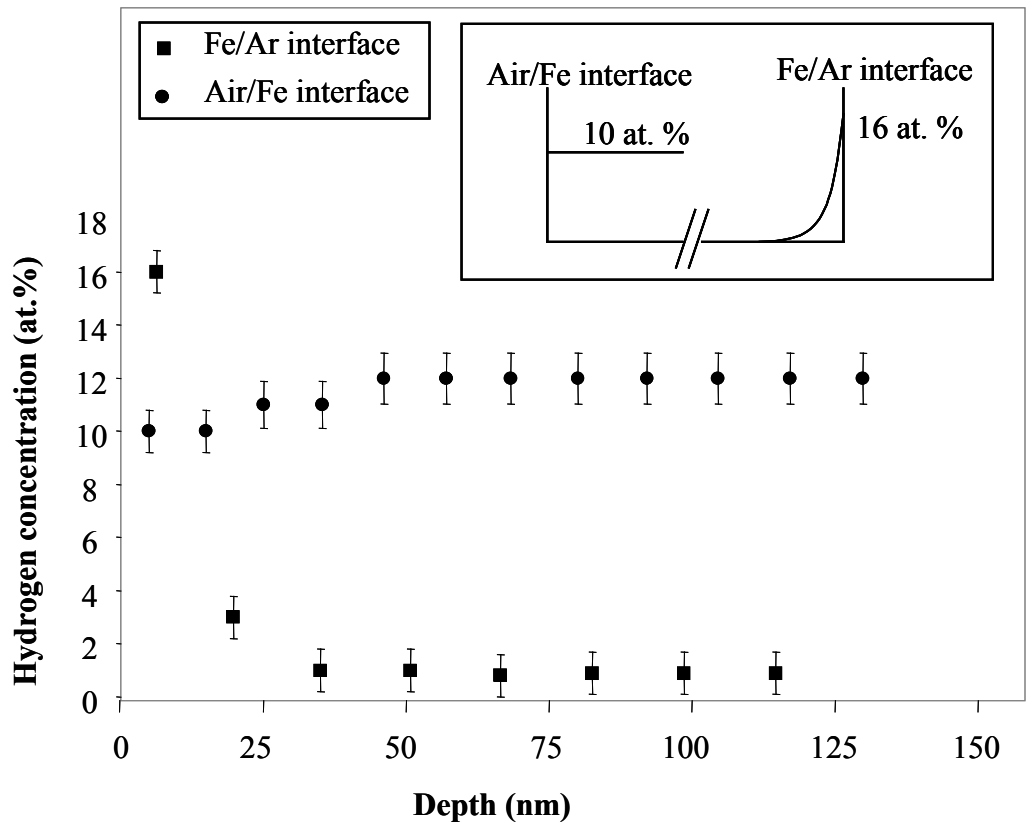

Figure 7